\documentclass[hyper,12pt]{article}    

\usepackage{bm,amsmath,amssymb,latexsym,graphicx,euscript,multirow,color,url,verbatim}
\usepackage{epsf}
\usepackage[sort&compress,numbers]{natbib}
\usepackage{hyperref}
\usepackage{enumerate}

\usepackage{amsmath}
\usepackage{amssymb}
\usepackage{amscd}
\usepackage{amsthm}

\long\def\symbolfootnote[#1]#2{\begingroup%
\def\thefootnote{\fnsymbol{footnote}}\footnote[#1]{#2}\endgroup}

\newcommand{\Jdp}{\mathcal{J}{\kern-.27em d}}
\newcommand{\Adp}{\mathcal{A}{\kern-.18em d}}

\newcommand{\bps}[1]{$1/{#1}$-BPS}

\DeclareMathOperator{\SU}{SU}

\newcommand{\zbra}[1]{\Bigl\langle\relax{\kern-.4em}\Bigl\langle#1\Bigr\vert}
\newcommand{\zinnp}[2]{\Bigl\langle\relax{\kern-.4em}\Bigl\langle #1\Bigm\vert #2\Bigr\rangle}            
\newcommand{\zbok}[3]{\Bigl\langle\relax{\kern-.4em}\Bigl\langle #1\Bigl\vert #2\Bigr\vert#3\Bigr\rangle} 
\newcommand{\nn}{\notag}

\def\half{\frac{1}{2}}
\def\s5{\textbf{S}^5}

\begin{document}

\begin{titlepage}

\begin{flushright}
\small{WIS/11/11-DEC-DPPA}
\end{flushright}

\vspace{0.5cm}
\begin{center}
\Large\bf
Classical probes for $1/16$ SUSY operators
\end{center}

\vspace{0.2cm}
\begin{center}
{Matan Field\symbolfootnote[1]{matan.field@weizmann.ac.il}}\\
\vspace{0.6cm}
\small \textit{Department of Particle Physics
and Astrophysics,\\Weizmann Institute of Science, Rehovot 76100, Israel}
\end{center}

\vspace{0.5cm}
\begin{abstract}
\vspace{0.2cm}
\noindent

We consider 1/16 SUSY solutions in AdS/CFT. On the gravity side, Gutowski and Reall showed them to be charged, rotating black holes in $AdS_5$. On the CFT side, an initial construction for 1/16 SUSY operators in $\mathcal{N}=4$ SYM has been suggested by Berkooz et al., with a Fermi-sea operator describing the extremal state. In this work we analyze particle trajectories in the 1/16 SUSY black hole background, and show the analysis to be sensitive to the Fermi-level of the suggested operator in the CFT.

\end{abstract}
\vfil

\end{titlepage}

\tableofcontents
\section{Introduction} \setcounter{equation}{0}
By providing a non-perturbative definition of quantum gravity, the AdS/CFT correspondence  \cite{Maldacena:1997re} allows, in principle, for a complete quantum description of black holes. In practice, detailed computations are possible only in supersymmetric cases. Unfortunately, there are no genuine black hole solutions, with a macroscopic horizon, in the 1/2,~1/4, or 1/8-BPS sectors.
 In the 1/16-BPS sector, supersymmetric asymptotically anti-de Sitter black holes were constructed in \cite{Gutowski:2004ez,Gutowski:2004yv,Chong:2005da,hep-th/0601156}, as solutions to five-dimensional gauged supergravity. 
 These supersymmetric black holes are charged and rotating, and, surprisingly, they constitute only a co dimension one surface of the space of extremal black holes.\footnote{Recently, a suggestion for a possible completion of this space was put forward in \cite{arXiv:1005.1287}.}

The quantum description of these black holes is possible via the CFT, which is strongly coupled in the regime of interest, and, for that, supersymmetry  might be of use. A BPS operator sits in a short multiplet, and its dimension is fixed by the BPS condition. The size of a multiplet cannot be deformed when continuously changing the parameters of a theory (specifically the couplings), and accordingly the dimension of such operators is protected against quantum correction. A caveat occurs if two short multiplets are joined together to form a longer one \cite{Dolan:2002zh}, but this is not always possible \cite{Dolan:2002zh,Kinney:2005ej}. When short multiplets cannot join to form a longer multiplet, the BPS operator is called protected and its dimension is un-renormalized. In this case it is kept BPS in the strong coupling regime, if it is computed to be so at weak coupling, and certain results can be extrapolated from weak to strong coupling. There is no construction of exactly 1/16-BPS operators in $d=4,~\mathcal{N}=4$ SYM, so far, and in particular it is not known whether these operators (or some of them) are protected. The identification of the 1/16-BPS operators will allow for a study of black holes from a microscopic point of view, and for example, with this knowledge at hand one may be able of understanding the appearance of classical horizons from the classical limit of quantum states. This is the long term motivation for our study.

Some first steps towards understanding the \bps{16} sector of $\mathcal{N}=4$ SYM have been taken in \cite{Berkooz:2006wc}, in the context of AdS/CFT. The large black-hole limit, in which the dual black hole mass is dominated by angular momenta, was explored, and a suggested construction for highest weight primary operators was put forward. The constructed operators satisfy the 1/16-BPS formula, at tree-level, and were conjectured in \cite{Berkooz:2006wc} to be primaries, up to addition of descendants.
The main ingredient for the 1/16-BPS construction was a Fermi-sea of $\mathcal{N}=4$ gauginos, which was filled up (in a gauge invariant way) by their angular momenta, up to some level. To the gauginos Fermi-sea an additional family of bosonic structures was added (in order to satisfy the exact 1/16-BPS equation), and altogether the final class of operators reproduced the correct entropy and angular-momentum-to-charge scaling, as predicted by the corresponding black hole solution, up to order one coefficients. It was explained in \cite{Berkooz:2006wc} that their construction can be generalized with some other bosonic structures over the Fermi-sea `core', to give additional 1/16-BPS operators, and thus the mismatch between the numerical coefficients was not a surprise.

In the construction of \cite{Berkooz:2006wc} the Fermi-sea structure was essential for the operator to be supersymmetric. The action of a specific supercharge on any `parton' in the Fermi-sea was shown to annihilate the parton into two other partons, which were already included in the sea, thus leading to the annihilation of the whole Fermi-sea operator by that supercharge. Later, the `naked' Fermi-sea operators (with no additional structure) were also shown to be weakly renormalized to 2-loop order  \cite{arXiv:0807.0559}, with their anomalous dimension suppressed by the Fermi-level. We expect that this Fermi-sea structure would be essential for any improvement of the model of \cite{Berkooz:2006wc} (that will give the exact counting, for example), and accordingly we look for a robust prediction of these Fermi-sea operators on the theory. Interestingly, such operators have a very sharp signal, which is the level of the Fermi-surface; the existence of the Fermi-surface is indeed insensitive to modifications of these operators in the bosonic sector. If the Fermi-sea model for minimally supersymmetric operators is correct, then a similar indication should be found on the gravity side as well.

Having this expectation in mind we consider in this work the 1/16-BPS black hole, and study trajectories of classical test-particle whirling around it. We analyze these geodesics as a function of the particle's energy, charge and angular momentum (per unit of mass). We find this analysis to be sensitive to the Fermi-level of \cite{Berkooz:2006wc}.

The paper is organized as follows. We begin in section \ref{sec:1/16BH} by summarizing some facts about 1/16-BPS black holes \cite{Gutowski:2004ez}. In section \ref{sec:1/16operators} we summarize the suggested initial construction of 1/16-BPS operators in $\mathcal{N}=4$ SYM \cite{Berkooz:2006wc}. In section \ref{sec:Schwarz} we warm up by working out a simpler example, analyzing the geodesic trajectories of a classical test-particle around the Schwarzschild black hole in $AdS_5$. In section \ref{sec:Fermi-surface}, in which we present our main result, we work out a geodesic analysis in the metric of the 1/16-BPS black hole of \cite{Gutowski:2004ez},
and find a clear probe for the Fermi-level of the suggested operators of \cite{Berkooz:2006wc}.

\section{1/16 SUSY black holes}\label{sec:1/16BH} \setcounter{equation}{0}
  The first supersymmetric, asymptotically $AdS_5$, black hole solutions were constructed in \cite{Gutowski:2004ez}. These formed a 1-parameter family of solutions of minimal five-dimensional gauged supergravity.\footnote{These 1/16-BPS solutions of the minimal gauged supergravity theory, with a single Abelian gauge field, can be uplifted to give BPS solutions of type IIB supergravity on $AdS_5\times S^5$ \cite{Chamblin:1999tk}. They can also be embedded into an $\mathcal{N}=1$ gauged supergravity with $U(1)^3$ gauge symmetry; the solution of \cite{Gutowski:2004ez} is then lifted to a solution of three equal charges.} These black holes were found to be charged and rotating, and to preserve 1/16 of the supersymmetries. Later on these solution were generalized in \cite{Gutowski:2004yv,Chong:2005da,hep-th/0601156}, but we will focus here on the special case of \cite{Gutowski:2004ez}, with their notations.
 \subsection{The background}
 The black hole solution of \cite{Gutowski:2004ez} carries one charge and a single angular momentum on the $S^3$ in $AdS_5$, say $J\equiv\half(J_1+J_2)$. The second angular momentum was put to zero, $\bar{J}\equiv\half(J_1-J_2)=0$, with $J_1,J_2$ the angular momenta on two orthogonal 2-planes. Even though the BPS formula should only reduce, generically, one of the remaining free parameters, M, J and Q, the solutions of \cite{Gutowski:2004ez} were found to contain, surprisingly, only a one parameter family of solutions. These black holes solutions were found to have the following metric,
\begin{equation}\label{eq:1/16_metric1}
ds^2=-f^2(R)dt^2-2f^2(R)\Psi(R) dt\sigma_L^3+U(R)^{-1}dR^2+\frac{R^2}{4}\Big[(\sigma_L^1)^2+(\sigma_L^2)^2+\Lambda(R)(\sigma_L^3)^2\Big]~,
\end{equation}
 with
 \begin{align}\label{eq:Rfunctions}
 U(R) & =\left(1-\frac{R_0^2}{R^2}\right)^2\left(1+\frac{2R_0^2}{l^2}+\frac{R^2}{l^2}\right)~~,~~ \Lambda(R)=1+\frac{R_0^6}{l^2R^4}-\frac{R_0^8}{4l^2R^6} ~,\nn\\
 f(R)&=1-\frac{R_0^2}{R^2}~~,~~\Psi(R)=\frac{-\epsilon R^2}{2l}\left(1+\frac{2R_0^2}{R^2}+\frac{3R_0^4}{2R^2\left(R^2-R_0^2\right)}\right)~,
 \end{align}
and $\epsilon=\pm1$ indicates the direction of rotation. Finally, $l$ is the radius of the AdS, and the $\sigma$'s used above are right invariant 1-forms on $SU(2)$, which can be expressed in terms of Euler angles ($\theta,\psi,\phi$) as:
 \begin{align}
 \sigma_L^1 & =\sin\phi d\theta-\cos\phi \sin\theta d\psi~,  \nn \\
 \sigma_L^2 & =\cos\phi d\theta+\sin\phi \sin\theta d\psi~,  \nn \\
 \sigma_L^3 & =d\phi+\cos\theta d\psi~. \label{eq:left inv forms}
 \end{align}
  The Maxwell potential is
  \begin{equation}
 A=\frac{\sqrt{3}}{2}\Big[f(R)dt+V(R)\sigma_L^3\Big]~. \label{eq:potential}
 \end{equation}
  The black hole horizon is situated at $R=R_0$ and is not spherical, but instead has the shape of a squashed sphere. The original isometry group of $AdS_5$ is broken by the black hole angular momentum into a subgroup $R_t \times U(1)_L \times SU(2)_R$. Recall, however, that far away from the horizon, since the metric approaches those of $AdS_5$, the spherical symmetry $SU(2)_L \times SU(2)_R$ is restored.

 This coordinate system is convenient to work with, but it doesn't have the standard $AdS_5$ asymptotics, so comparisons with the CFT needs to be considered carefully. To make the $AdS_5$ asymptotics manifest, a change of coordinates is introduced,
 \begin{align}\label{eq:coord_trans}
 \phi'=\phi+2\epsilon t~.
 \end{align}
  The metric is then
  \begin{equation}\label{eq:1/16_metric2}
ds^2=-U(R)\Lambda(R)^{-1}dt^2+U(R)^{-1}dR^2+
\frac{R^2}{4}\Big[(\sigma_L^1)^2+(\sigma_L^2)^2+\Lambda(R)\left(\sigma_L^3-\Omega(R)dt\right)^2\Big]~,
\end{equation}
and the Maxwell potential is
  \begin{equation}
 A=\frac{\sqrt{3}}{2}\Big[h(R)dt+V(R)\sigma_L^3\Big]~, \label{eq:potential}
 \end{equation}
 with
 \begin{align}\label{eq:Rfunctions}
 \Omega(R)&=\frac{2\epsilon}{l\Lambda(R)}\left[\left(\frac{3}{2}+\frac{R_0^2}{l^2}\right)\frac{R_0^4}{R^4}
 -\left(\half+\frac{R_0^2}{4l^2}\right)\frac{R_0^6}{R^6}\right] ~,\nn\\
 h(R)&=1-\frac{R_0^2}{R^2}-\frac{R_0^2}{2R^4}~.
 \end{align}
In section \ref{sec:Fermi-surface} we present our main computation in the former coordinate system (which makes things simpler), and will then take care for the needed translation from the gravity conserved charges to the CFT ones, due to this coordinate transformation. We mention, however, that in order to verify our computation (which is rather technical) we have proceeded with both coordinate systems and obtained equivalent results.
 \subsection{Properties of the solution} \setcounter{equation}{0}
 The mass, angular momentum and charge of the black hole are:
 \begin{align}\label{eq:MJQ}
M&=\frac{3\pi R_0^2}{4G}\bigg(1+\frac{3R_0^2}{2l^2}+\frac{2R_0^4}{3l^4}\bigg)\approx \frac{\pi R_0^6}{2G l^4}\nn\\
J&=\frac{3\epsilon \pi R_0^4}{8Gl}\bigg(1+\frac{2R_0^2}{3l^2}\bigg)\approx \frac{\epsilon \pi R_0^6}{4Gl^3}\nn\\
Q&=\frac{\sqrt{3}\pi R_0^2}{2G}\bigg(1+\frac{R_0^2}{2l^2}\bigg)\approx \frac{\sqrt{3}\pi R_0^4}{4Gl^2}~.
\end{align}
 These saturate the 1/16-BPS bound,\footnote{The generic 1/16 BPS bound is $M=\frac{2|J|}{l}+Q_1+Q_2+Q_3$, and here we define $Q_1=Q_2=Q_3\equiv \frac{Q}{2\sqrt{3}}$, consistently with \cite{Berkooz:2006wc}.}
 \begin{equation}
 M=\frac{2|J|}{l}+\frac{\sqrt{3}}{2}Q~.
 \end{equation}
 Everywhere in this work the classical gravity limit is being assumed, that is the large N and large 't Hooft coupling limit in the field theory. From now on we also set $l=1$.
 We are also working in the very large mass limit $R_0^2>>1$, and in the limit where $\bar J=0$, in which the charge and angular momentum of the black hole were shown \cite{Chong:2005da,Berkooz:2006wc} to satisfy the following relation,
 \begin{align}\label{eq:JQscaling}
 \frac{J}{N^2}=\sqrt{2}\left(\frac{Q}{\sqrt{3}N^2}\right)^{3/2}~.
 \end{align}
 Comparing this with the form of \cite{Gutowski:2004ez} \eqref{eq:MJQ}, we see the black hole charge and angular momentum can also be written as
 \begin{align}\label{eq:JQ2}
 J=\frac{N^2x_0^3}{2}~~,~~Q=\frac{\sqrt{3}N^2x_0^2}{2}~~,
 \end{align}
 where we have also defined $x_0\equiv R_0^2$ for later convenience. This form will be most natural for
 comparison with the field theory predictions.
 The scaling relation \eqref{eq:JQscaling} was obtained in \cite{Berkooz:2006wc} by the Fermi-sea construction, but with
 $\sqrt{2}/3$ replacing $\sqrt{2}$.\footnote{This is for the `naked' Fermi-sea operators, before the addition of the bosonic structure. Adding the latter modifies this coefficient, but not enough to compensate for the factor of $3$ mismatch.} One possibility for this mismatch is that the angular momentum of the Fermi-sea operator is smaller than that of the black hole by a factor of $3$. We will see in the sequel that this is exactly the option that is consistent with our results.

\section{A suggested construction for 1/16 SUSY operators at weak coupling}\label{sec:1/16operators} \setcounter{equation}{0}

The primary motivation for the current study is the identification of all 1/16-BPS operators in $\mathcal{N}=4$ SYM. A starting point in this program was taken in \cite{Berkooz:2006wc},
 where a family of operators that satisfy the 1/16-BPS formula \cite{Dolan:2002zh}, at tree level, was constructed. The operators were shown to satisfy the correct angular momentum to R-charge scaling relation \eqref{eq:JQscaling}, as found in the gravity solution, up to an order one coefficient. While these operators were not fully proven to be primaries, some arguments indicating they cannot be written as descendants of primaries were sketched. Thus, \cite{Berkooz:2006wc} made the conjecture that these operators are primaries, up to the addition of a descendant. When counting them, they have been found to give the correct Bekenstein-Hawking entropy, again up to an order one coefficient. In this section we briefly review the essential ingredients in the work of \cite{Berkooz:2006wc}, trying to emphasize the Fermi-sea element of the construction. To keep the flow of the paper we try to avoid technical details as much as possible, and we send the reader to \cite{Berkooz:2006wc} for the full details of their work.

\subsection{$\mathcal{N}=4$ SYM and 1/16-BPS representation} \setcounter{equation}{0}
The field content of $\mathcal{N}=4$ SYM includes the chiral and anti-chiral parts of the gauge boson field strength ($F_{\alpha \beta},~\bar{F}_{\dot{\alpha} \dot{\beta}}$), the four Weyl spinor gauginos ($\lambda_{i\alpha}~,\bar{\lambda^i_{\alpha}}$), and six real scalars, which can also be packed within six complex scalars ($M_{ij}$), obeying the reality condition $(M_{ij})^{\dagger}=\bar M^{ij}\equiv\half\epsilon^{ijkl}M_{kl}$. Here, undotted $(\alpha)$, dotted $(\dot{\alpha})$ Greek indices and Latin $(i)$ indices stand for $SU(2)_L \times SU(2)_R \times SU(4)$ symmetry indices, respectively.
The theory is invariant under the $\mathcal{N}=4$ supersymmetry transformations of the schematic form
\begin{align}\label{eq:SUSY trans}
\delta M \sim\lambda + \bar \lambda~,~~\delta \lambda\sim F  +iDM -i[M,\bar M]~,~~\delta F \sim iD\lambda-iD \bar \lambda  ~,
\end{align}
where $D$ stands for the covariant derivative.
This theory also admits a bosonic symmetry, including the conformal group in 4D Minkowski space $SO(2,4)$ and the $SU(4)$ R-symmetry.

The $\mathcal{N}=4$ superconformal representations are classified by six numbers, the conformal dimension ($\Delta$),
the $SU(2)_L \times SU(2)_R$ angular momenta on the $S^3$ in $AdS_5$  ($J,\bar J$), and the Dynkin labels of the $SU(4)$ representation ($[k,\,p,\,q]$). In (semi-)short representations, the conformal dimension is determined by the other numbers, and the short representation is denoted by
\begin{equation*}
  [k,\,p,\,q]_{J,\,\bar J}~.
\end{equation*}
The relation of the Dynkin labels to the $SU(4)$ charges which were written in the gravity is:
\begin{equation}
    Q_1=\frac{k+2p+q}{2 l}~,~~Q_2=\frac{k+q}{2 l}~,~~ Q_3=\frac{k-q}{2 l}~.
 \label{eq:rchargedict}
\end{equation}
Our case in the gravity \cite{Berkooz:2006wc} is the one with $Q_1=Q_2=Q_3\equiv \frac{Q}{2\sqrt{3}}$, which corresponds here to $p=q=0,~k=\frac{Ql}{\sqrt{3}}\equiv \mathcal{Q}$.

In \cite{Dolan:2002zh} a detailed analysis of short and semi-short representations
of the $\mathcal{N}=4$ superconformal group was carried out. In particular, a 1/16-BPS representation (denoted $c^{1/4}$) was found; In this representation, the superconformal primary operator, in addition to be annihilated by all the conformal supercharges, is also annihilated by a single additional combination out of the sixteen poincar\'{e} supercharges,
\begin{equation}\label{eq:1/16cond}
  |k,~p,~q;~J~,\bar J\rangle~\in~c^{1/4}~~\Leftrightarrow~~ \left(Q^1_2 - \frac{1}{2J+1}~J_-~Q^1_1\right)~|k,~p,~q;~J,~\bar J\rangle  = 0~,
\end{equation}
where $J_-$ stands for the lowering operator in $\SU(2)_L$ and
$Q^i_\alpha$ are the supercharge generators. Further, the conformal dimension $\Delta$ of these superconformal primary operators was found to obey the 1/16-BPS formula:
\begin{gather}\label{eq:BPS CFT}
    \Delta^{(c^{1/4})}=2+2J+\frac{3}{2}k+p+\frac{1}{2}q~.
\end{gather}
When relating this to the BPS formula in the gravity (using the $Q_i\leftrightarrow [k,p,q]$ and $\Delta \simeq Ml$ dictionary), we discern a difference of factor 2. However, this additive factor is unobservable in the regime we are working, where the charges are generically very large to ensure a reliable classical spacetime description.
\subsection{Fermi-sea operators} \setcounter{equation}{0}
 One of the key observation in \cite{Berkooz:2006wc} was to identify some remarkable properties of the fermionic operator, denoted by $A^{(I)}$, that is built up from the action of $I$ covariant derivatives on a gaugino. Schematically it is written as
\begin{align}
  A^{(I)\,1}\sim D^I \bar \lambda^1~,    \label{eq:A-def}
\end{align}
where $A^{(I)}$ has angular momenta $J=I/2,$ and $\bar J=(I+1)/2$. This operator features three important properties. The first is its supersymmetry transformation, schematically written,
\begin{align}\label{eq_Q_of_A}
\left\{Q^1,A^{(I)1}\right\}\sim \sum_{m=1}^{I}\left\{A^{(m-1)1},A^{(I-m)1}\right\}~.
\end{align}
This means that the supersymmetry action on $A$ splits it into two $A$'s, which always have lower angular momenta than the original one. Secondly, its dimension satisfies the 1/16-BPS formula, apart from its `global' part of an additive factor of $2$,
\begin{align}
\Delta[A^{(I)}]= 2J+\frac{3}{2}k[A^{(I)}]+p[A^{(I)}]+\frac{1}{2}q[A^{(I)}]=\Delta^{(c^{1/4})}[A^{(I)}]-2~.
\end{align}
Thirdly, while its R-charge is fixed, its angular momenta grow linearly with $I$. The authors of \cite{Berkooz:2006wc} have used the above first property to construct a Fermi-sea out of these $A$-operators, by multiplying all $A$-operators up to some left angular momentum level $K$. This is done in a gauge invariant way,
 \begin{align}\label{eq:Jdet}
 \mathcal{J}_{\mbox{\tiny{Fermi-sea}}}^{(K)}\equiv\prod_{I=0}^K \prod_{m=0}^{I+1} \mbox{Jdet}\Big[\left(\bar J_-\right)^m A^{(I)1}_{hw}\big]~,
 \end{align}
where
\begin{align}
\mbox{Jdet}[X]=\epsilon_{a_1a_2...a_g}X^{a_1}X^{a_2}...X^{a_g}~~,~~X=\sum_{a=1}^{N^2-1} X^a T^a~,
\end{align}
and $A_{hw}$ is an $SU(2)_L\times SU(2)_R$ highest weight operator.
Notice that $\bar J_-$ generates also the multiplication of the entire $SU(2)_R$ multiplet, and thus
 makes the right angular momentum to vanish, $\bar J=0$. The fermionic building block of the Fermi-sea operator, the $A$'s, are referred to as `partons'.
 The angular momentum and charge of these Fermi-sea operators were also computed at the large $N$ and large $K$ limit \cite{Berkooz:2006wc}:
 \begin{align}
 J=\frac{N^2 K^3}{6}~~,~~Q= \frac{\sqrt{3}N^2 K^2}{2}~.
 \end{align}
 This can be seen to almost satisfy \eqref{eq:JQscaling}, up to a factor of $3$. We see that the relation of the Fermi-level and the black-hole parameter is
  \begin{align}
 K\sim x_0~,
 \end{align}
 up to an order one coefficient. \emph{If the previous mismatch from \eqref{eq:JQscaling} is completely due to the value of $J$}, than we find that  simply $K=x_0$.
 We will find this possibility to be in a very good agreement with our results. The Fermi level of left angular momentum is then $x_0/2$.

 The Fermi-sea operator enjoys some desired features, as well as some repairable faults:
\begin{itemize}
\item It is annihilated by two supercharges. The relevant supercharges split each fermionic parton into two new fermionic partons which are already in the Fermi-sea, and thus by the Pauli exclusion principle the whole Fermi-sea operator is annihilated by those supercharges. This still calls for a remedy, as the final operator needs to be annihilated by a single supercharge only.
\item The Fermi-sea operator satisfies the 1/16-BPS formula, apart from the factor $2$ global part, at weak coupling, by just adding the quantum numbers of all partons in the sea. This still needs to be added an operator with dimension $2$.
\item In the large $K$ limit, which is equivalent to the large mass limit in the supergravity, the Fermi-sea operator has an angular momentum which is much larger than its R-charge, and it was shown in \cite{Berkooz:2006wc} to exactly reproduce the correct scaling behavior \eqref{eq:JQscaling}, up to an order one multiplicative factor.
\end{itemize}
Another fault of the `naked' Fermi-sea operator is that it is so far unique, for a given total angular momentum, and still degeneracy needs to be introduced into this picture, in order to generate the desired entropy. It turned out \cite{Berkooz:2006wc} that by introducing additional bosonic structure (with tree-level dimension $2$), to be multiplied with the `naked' Fermi-sea operator, all above faults were apparently fixed. The 1/16-BPS formula was exactly satisfied; the degeneracies were introduced, and the computed entropy agreed with the supergravity prediction, again, up to an order one coefficient. Finally, the operator was argued to be a genuine primary, up to an addition of a descendent. Here we have described the case of $\bar J=0$, which is the case of our interest, but also the case of $J=\bar J$ has been worked out in \cite{Berkooz:2006wc}.

\section{A toy model: Schwarzschild black hole}\label{sec:Schwarz} \setcounter{equation}{0}
We consider a classical test particle in the $AdS_5$ Schwarzschild background. The rotational symmetry $SO(4)\simeq SU(2)_L \times SU(2)_R$ constrains the motion of the particle onto a plane, which can be described then by a single angular momentum $j$. The particle has also energy $E$  and a mass $m$ ($j$ and $E$ are per unit of mass if the particle is massive).
In this section we study the trajectories of the particle, classified by its constants-of-motion, according to whether or not it is falling into the black hole. It should be noted that while our particle either falls quickly or not at all, a real quantum particle has a wave function which is always leaking into the horizon. Our classical particle is an appropriate limit for the case of very large quantum numbers, and we understand a (non-) falling classical particle as one which is endowed with a very (slow) fast decaying wave function.

 \subsection{Geodesic analysis} \setcounter{equation}{0}
The metric of a Schwarzschild black hole in $AdS_5$ is the following \cite{Witten:1998zw}
\begin{equation}\label{eq:Sch-metric}
ds^2=-\bigg(1+r^2-\frac{\lambda}{r^2}\bigg)dt^2+\bigg(1+r^2-\frac{\lambda}{r^2}\bigg)^{-1} dr^2+r^2d\Omega^2 ~,
\end{equation}
where $\lambda=\frac{8}{3\pi} G_5 M$, and $d\Omega^2$ is the standard metric on the 3-sphere,
\begin{equation}\label{eq:3-sphere}
d\Omega^2=d\theta^2+\sin^2 \theta d\phi^2+\sin^2 \theta \sin^2 \phi d\psi^2.
\end{equation}
In the coordinates of \eqref{eq:3-sphere} we choose the plane of motion to be at $\theta=\phi=\frac{\pi}{2}$, so that the angular coordinate in the plane is $\psi$.
For every Killing vector $K\equiv K^M\partial_M$ that satisfies the Killing equation $\nabla_M K_N+\nabla_N K_M=0~$, where $\nabla$ is the general covariant derivative, there is a constant-of-motion for the particle's trajectory $C_K=g_{MN}K^M\dot X^N~.$
   For the Killing vectors $\partial_{\psi}$ and $\partial_t$ we find the energy and angular momentum:
\begin{align}\label{eq:EJ definition}
E =&~\bigg(1+r^2-\frac{\lambda}{r^2}\bigg)\frac{dt}{d\tau} \\
j =&~ r^2\frac{d\psi}{d\tau}~.
\end{align}
Writing the equations-of-motion for a massive particle
\begin{align}
g_{MN}\dot{X}^{M}\dot{X}^{N}=-1~,
 \end{align}
 and putting in the definitions for $E$ and $j$, we find for the radial coordinate an effective one-dimensional problem
\begin{align}
\dot{r}^2+V(r)=\varepsilon ~.
\end{align}
The effective potential and effective energy are:
\begin{equation}\label{eq:eff potential}
V(r)=\left(r^2  +\frac{j^2-\lambda}{r^2} -\frac{\lambda j^2}{r^4}\right)~~,~~
\varepsilon=E^2-j^2-1~,
\end{equation}

 To understand whether or not the real particle is falling into the black hole, we need to analyze the effective potential structure to decide whether the effective particle (with its effective energy  $\varepsilon$) will roll down to $ r=r_h$, or not. We work in the large black hole limit ($\lambda^{1/2}>>1$), in which the horizon is situated at $r_h=\lambda^{1/4}$.
We define
$$ V_1\equiv  r^2 ~,~ V_2\equiv \frac{j^2-\lambda}{ r^2}~,~ V_3\equiv  -\frac{j^2 \lambda }{ r^4}~,$$
such that $V=V_1+V_2+V_3$. For the effective particle not to fall all the way to $ r=r_h$ we need to have a minimum $r_-$ and a maximum $r_+$ ($r_-> r_+$). To have an extremum we need
$$\half V'(r)= r - \frac{j^2-\lambda}{r^3} + \frac{2j^2\lambda}{r^5}=0 ~.$$
First we see that $V_1'$ and $V_3'$ are always positive, thus we must demand $V'_2<0$ in order to have a minimum. This implies $j^2>\lambda$. Looking at the generic graphs of $V_{1,2,3}$ separately, we see that $V_1+V_2$ 'generate' a minimum which is 'shifted to the left' by $V_3$, while $V_2+V_3$ 'generate' a maximum which is 'shifted to the right' by $V_2$. Analyzing this accurately we find
\begin{align}\label{rhos_ineq}
r_h<\left(\frac{2j^2\lambda}{j^2-\lambda}\right)^{\half}< r_+< r_-<\left(j^2-\lambda\right)^{\frac{1}{4}} ~,
\end{align}
from which we get
\begin{align}
j^6-4\lambda^2 j^4>3\lambda j^4-3\lambda^2 j^2 + \lambda^3 > 0 ~,
\end{align}
 and derive the inequality
\begin{equation}\label{eq:first bound}
j^2>4\lambda^2~.
\end{equation}
From this we see that we can approximate $j^2-\lambda\approx j^2$
when $\lambda>>1$. Thus, when looking back at the potential, and defining
$y= r/\lambda^{\half}~,~ k=j^2/\lambda^{2}~,$ we get a simpler problem:
\begin{align}\label{final equation}
&\dot{y}^2+V(y)=\varepsilon~,~~V(y)=y^2 +\frac{k}{y^2} -\frac{k}{y^4}~,~~\varepsilon=\frac{E^2-\lambda^2 k}{\lambda}~.
\end{align}
For the particle not to fall we must have extrema of $V(y)$ at $y>y_h=\lambda^{-\frac{1}{4}}$,
$$\half V'(y)=y-\frac{k}{y^3}+\frac{2k}{y^5}=0~.$$
After replacing $x=y^2$ this amounts to
\begin{equation}\label{eq:basic equation}
f(x)=x^3-kx+2k=0~~.
\end{equation}
\\
Analyzing analytically this polynomial, we find that it has positive zeros if and only if $k>27$ and both of them always lies in front of the horizon. Reducing it back to the original variables it comes to
\begin{equation}\label{eq:jbound}
j>\sqrt{27} \lambda~.
\end{equation}
For the particle not to fall, we also need to demand a bound on the effective energy,
\begin{align}
V(x_-)<\varepsilon<V(x_+)~,
\end{align}
and we find that this energy window narrows in the large $\lambda$ limit, such that finally
\begin{align}\label{eq:Ebound}
E\sim j~.
\end{align}

We have found sharp bounds on the particle's energy \eqref{eq:Ebound} and angular momentum \eqref{eq:jbound} not to fall to the horizon. This can be interpreted as a robust probe for the black hole solution, and as such, a holographic construction of this black hole should be able of reproducing the same sharp signals.

\section{Probing the Fermi-surface}\label{sec:Fermi-surface} \setcounter{equation}{0}
In this section we will perturb the 1/16-BPS black hole solution by considering a classical particle moving around it. Similarly to the previous section, we will reduce the problem to a one-dimensional effective problem for the radial direction, and will analyze the particle's trajectory according to its charge, energy and angular momentum. We will see that the analysis is very much sensitive to the level of the Fermi-surface, which is predicted from the suggested construction of 1/16-BPS operators in the CFT. As the details of the computations are very technical, and not very illuminating, we will only quote here our steps and results.
\subsection{Symmetry and constants-of-motion}
The black hole metric and Maxwell potential are invariant under the $SU(2)_R$ right part of the original rotational symmetry, and in addition under an Abelian subgroup of the left part $U(1)\subset SU(2)_L$, which are rotations of $\phi$. We also have a symmetry for translations in time.
 Together it is $R_t\times U(1)_L\times SU(2)_R ~$, with the 5 Killing vectors:
\begin{align}\label{eq:1/16_Killing_vectors}
 \xi_1^R & =  -\cot\theta \cos\psi \partial_{\psi} - \sin\psi \partial_{\theta} + \frac{\cos\psi}{\sin\theta}\partial_{\phi}~,\nn \\
 \xi_2^R & =  -\cot\theta \sin\psi \partial_{\psi} + \cos\psi \partial_{\theta} + \frac{\sin\psi}{\sin\theta}\partial_{\phi}~,\nn\\
 \xi_3^R & =  \partial_{\psi}~,\nn\\
 \xi_3^L & =  \partial_{\phi}~, \nn\\
 \xi_t & =  -\partial_t ~~,
\end{align}
about which more details can be found in Appendix A of \cite{Gauntlett:1995fu}.
In the presence of an electromagnetic potential, the constants-of-motion associated with any Killing vector $\xi=\xi^M\partial_M$, are
\begin{align}\label{eq:define_conserved}
C_{\xi}= \xi^M\left(g_{MN}\dot{X}^{N}+qA_{M}\right)~,
\end{align}
where $q$ is the particle's charge per unit of mass. In this case the Killing vector needs to satisfy a modified Killing equation, depended both on the metric and on the electromagnetic potential. In the simplest case it satisfies both parts separately,
\begin{align}\label{eq:Killing2}
0=&\partial_{P}V^{M}g_{MQ}+\partial_Q V^{M}g_{MP}+V^M\partial_M g_{PQ}~,\nn\\
0=&\partial_P V^M A_M + V^M\partial_M  A_P~,
\end{align}
as it is the case for the above Killing vectors.
Using \eqref{eq:1/16_metric1} and \eqref{eq:1/16_Killing_vectors} in \eqref{eq:define_conserved}
 we find the constants-of-motion:
 \begin{align}\label{eq:COM}
j_1^R & = \frac{1}{4}\left[\left(4\rho V + R^2\Lambda\dot{\phi}-4f^2\Psi\dot{t}+
R^2(\Lambda-1)\dot{\psi}\cos\theta\right)\cos\psi \sin\theta-R^2\dot{\theta}\sin\psi\right]~, \nn\\
j_1^R & = \frac{1}{4}\left[\left(4\rho V + R^2\Lambda\dot{\phi}-4f^2\Psi\dot{t}+
R^2(\Lambda-1)\dot{\psi}\cos\theta\right)\sin\psi \sin\theta+R^2\dot{\theta}\cos\psi\right]~, \nn\\
 j_3^R & = \frac{1}{4}\left[ \left(4\rho V+ R^2\Lambda\dot{\phi}-4f^2\Psi\dot{t}+R^2\Lambda\dot{\psi}\cos\theta\right)\cos\theta
 +R^2\dot{\psi}\sin^2\theta\right]~, \nn\\
 j^L_3 & =  \frac{R^2\Lambda}{4} \left(\dot{\phi}+\dot{\psi}\cos\theta\right)+\rho V-f^2\Psi\dot{t}~,\nn \\
 E & =  f^2\left(\dot{t}+\Psi(\dot{\phi}+\dot{\psi}\cos\theta)\right)-\rho f~,
\end{align}
where we have defined the modified charge $\rho = \sqrt{3}q/2$.
The above constants-of-motion satisfy the following relations:
\begin{align}\label{eq:charges relations}
 0=&j_2^R\cos\psi -j_1^R\sin\psi  -  \frac{R^2}{4}\dot{\theta}~,\nn\\
0=&j_3^R  -j_3^L\cos\theta -\frac{R^2}{4}\dot{\psi}\sin^2\theta~, \nn\\
  0=&j_3^L-  j_3^R\cos\theta - \left(j_2^R\sin\psi  +j_1^R\cos\psi\right)\sin\theta~,
\end{align}
and using the $SU(2)_R$ symmetry we set $j_1^R=j^R_2=0~,~j^{R,L}_3\equiv j_{R,L},$
to give
\begin{align}\label{eq:thetapsi}
\cos\theta=\frac{j_L}{j_R}~~,~~\dot{\psi}=\frac{4j_R}{R^2}~.
\end{align}
From \eqref{eq:COM} and \eqref{eq:thetapsi} we also find
\begin{align}\label{eq:phit}
  \dot{t}  &= -\frac{4f^2\Psi}{R^2U}\Big(E\Psi+j_L+\rho(f\Psi-V)\Big)+\frac{E+\rho f}{f^2} ~, \nn \\
  \dot{\phi} &= \frac{4f^2}{R^2U}\Big(E\Psi+ j_L +\rho(f\Psi-V)\Big)-\frac{4j_L}{R^2}~.
\end{align}
Notice that $j_L/j_R=\cos \theta$ implies that always $j_L\leq j_R$, which is indeed what we expect when we account for the restored symmetry at infinity.
%
%
The above constants-of-motion are the energy and angular momenta of the particle per unit of mass, from the bulk point of view. However, in order to conform with the CFT standard quantities we need to make the coordinate transformation \eqref{eq:coord_trans}, by which the standard $AdS_5$ asymptotic is obtained \cite{Gutowski:2004ez}. Taking \eqref{eq:coord_trans}, the corresponding Killing vectors transform as:
  \begin{align}
&\partial_{\phi'}=\partial_ {\phi}~,\nn\\
&\partial_{t'}=\partial_t-2\varepsilon\partial_{\phi}~,
\end{align}
and this results in a shift of the energy\footnote{Note that, as usual, the energy is defined to be the constant-of-motion of $-\partial_t$, so that it is positive.},
\begin{align}\label{eq:Ebulk}
E_{CFT}=E_{Bulk}+2\epsilon j_L~.
\end{align}
We will keep using below $E$ for $E_{Bulk}$, and take into account the transformation \eqref{eq:Ebulk}
at the end.

\subsection{The geodesic equation} \setcounter{equation}{0}
We put \eqref{eq:thetapsi} and \eqref{eq:phit} into the geodesic equation
$$g_{MN}\dot{X}^{M}\dot{X}^{N}=-1~,$$
 for a massive particle, and find the one-dimensional problem
 \begin{align}\label{eq:REOM}
 \dot{R}^2+V(R)=0~,
 \end{align}
 with the effective potential
 \begin{align}\label{eq:RPotential}
 V(R)=U\left(1+\frac{4\left(j_R^2-j_L^2\right)}{R^2}\right)-\frac{U}{f^2}\big(E+\rho f\big)^2
 +\frac{4f^2}{R^2}\Big(E\Psi+j_L+\rho(f\Psi-V)\Big)^2~.
 \end{align}
 Defining $x=R^2/R_0^2$ and $x_0=R_0^2$, with which
the horizon is at $x=1$ and our very large mass limit is $x_0>>1$,
 and using \eqref{eq:Rfunctions} in \eqref{eq:REOM} and \eqref{eq:RPotential}, our one-dimensional problem is
\begin{align}\label{eq:xEOM}
\dot x^2+W(x)=0~,
\end{align}
with
\begin{align}\label{eq:xPotential}
W(x)=w_2 x^2+w_1 x+w_0+\frac{w_{-1}}{x}+\frac{w_{-2}}{x^2}~,
\end{align}
and
\begin{align}
&w_2=4~,\nn\\
&w_1=\frac{4x}{x_0}\Big(4j_R^2-(E+\rho)^2\Big)~,\nn\\
&w_0=4\Big(\rho^2+E\rho-2\epsilon j_L \rho-3\Big)~,\nn\\
&w_{-1}=\frac{4}{x}\Big(-2\rho^2-2E\rho-E^2+4\epsilon j_L(E+\rho)-4j_L^2-\frac{12j_R^2}{x_0}+2\Big)~,\\
&w_{-2}=\frac{4}{x^2}\Big(\rho^2+E\rho+\frac{1}{4}E^2-\epsilon j_L(E+2\rho)+j_L^2+\frac{8j_R^2}{x_0}\Big)~.
\end{align}
In the above we have already ignored terms that are subleading in $1/x_0$, but without any assumptions about the order of magnitude of the parameters ($E,j_L,j_R,\rho$).

\subsection{Analysis} \setcounter{equation}{0}
\setcounter{equation}{0}
Our effective potential is a fairly complicated function of $R,E,j_L,j_R,\rho$ and $R_0$.
While we are looking for a specific (and robust) signal, the left angular momentum Fermi-Level, the potential encodes
much more information in it. This is very similar to the situation in collider physics, where the outcome of a collision is very complicated, and the search for a resonance needs to be dug out from the whole data.
  It turns useful to stretch the analogy, thus considering a specific `channel' for the analysis. Meaning,
we will fix most of the parameters in a natural way (put `cuts' on them), and leave only the desired parameter unfixed, which then will be used as a probe for the Fermi-level.

 In the `channel' we use we freeze all parameters of our test particle to be similar to the parameters of the partons in the Fermi-sea of \cite{Berkooz:2006wc}. This means fixing
 \begin{align}
 E_{CFT}=2j_L+3/2~,~q=\sqrt{3}~,~j_R=0~,\epsilon=1~.
 \end{align}
 In our bulk notations this means taking
 \begin{align}
 E=\rho=3/2~,~j_R=0~,\epsilon=1~,
 \end{align}
 where in the above the relation \eqref{eq:Ebulk} was used.
 Only $j_L$ is kept as a free parameter, and we are looking for the trace of the Fermi-level to appear as a sharp signal in the trajectories behavior, at the vicinity of $j_L\sim x_0$. Since $x_0>>1$, we find that in practice we can set $E\approx \rho\approx 0$, as the contribution of these small numbers will be dominated by the contribution from the large left angular momentum.

There are many possible checks for signals, or many things to look at; and for each case there is also the whole range of $j_L$ which needs to be examined. We have performed checks of many kinds, and thoroughly investigated the whole range of $j_L$, but we will present here only `positive' results, and just a few examples of them. The analysis is mostly numerical.

\begin{figure}[h]
\includegraphics[width=0.5\textwidth]{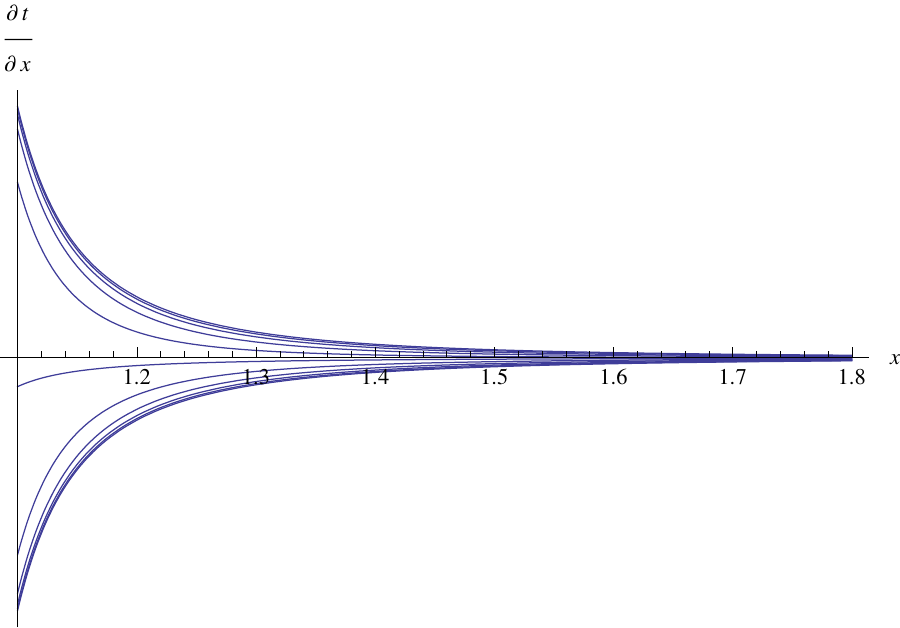}
\caption{Plots of $\frac{dt}{dx}$ for $j_L$ integer values in the range $45\leq j_L\leq 55$. The plots above the $x$-axis are for $j_L>50$. The geodesic starts at $x(\tau=0)=100$ with $\dot{x}(\tau=0)=0$.
The black hole parameter is $x_0=100$~.}
\label{fig:dtdx}
\end{figure}

\begin{figure}[h]
\includegraphics[width=0.5\textwidth]{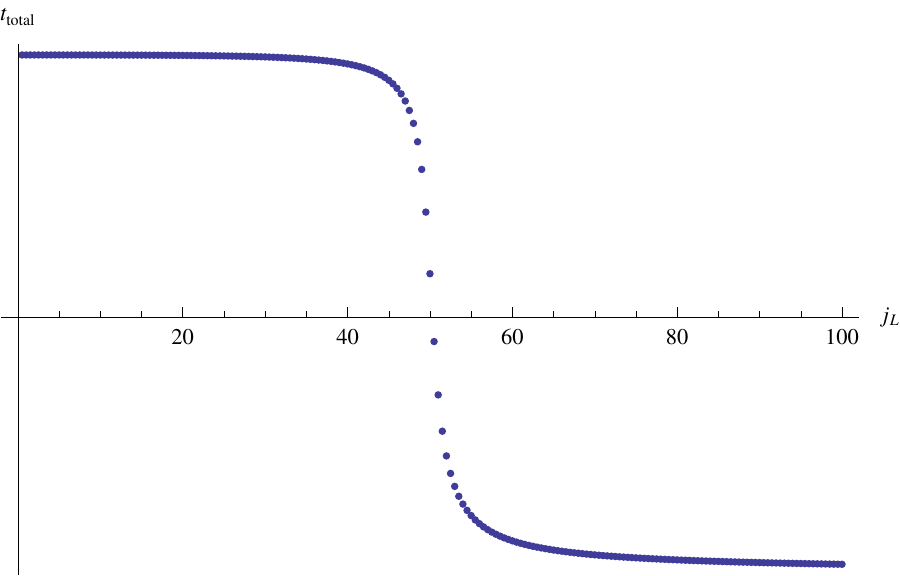}
\caption{Total falling global time for an infalling geodesic that start at $x(\tau=0)=100$ with $\dot{x}(\tau=0)=0$. $j_L$ is varied on integer values between $0$ and $100$. The integration is cut-off below $x=1.1~$. The black hole parameter is $x_0=100$.}
\label{fig:total global time}
\end{figure}

\begin{figure}[h]
\includegraphics[width=0.5\textwidth]{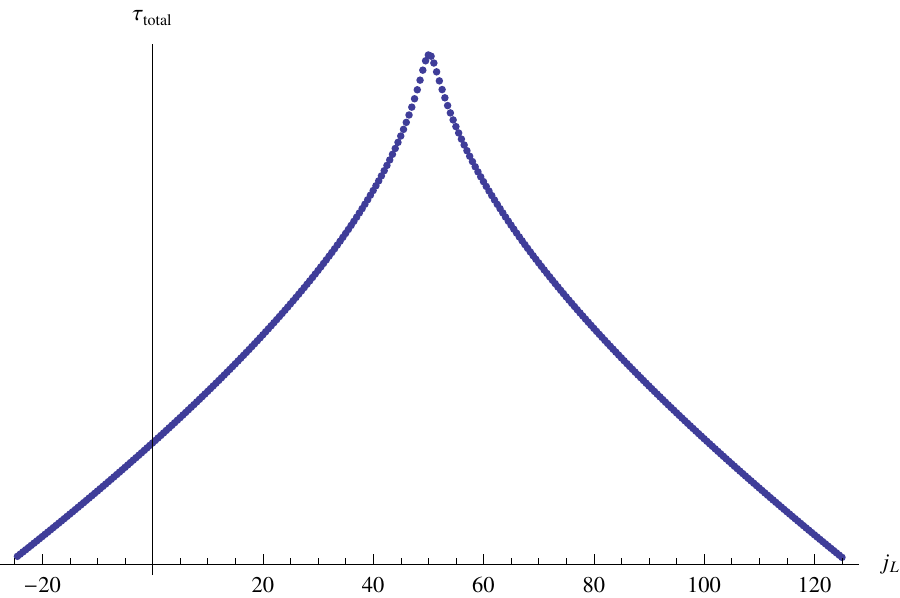}
\caption{Total falling self time for an infalling geodesic that start at $x(\tau=0)=100$ with $\dot{x}(\tau=0)=0$. $j_L$ is varied on integer values between $-25$ and $125$. Negative $j_L$ values are for angular momenta in the `other' direction, meaning $\epsilon=-1$. The integration is cut-off below $x=1.1$~. The black hole parameter is $x_0=100$.}
\label{fig:total self time}
\end{figure}

In all following examples we consider the black hole parameter to be $x_0=100$, and geodesics of falling particles that begins at $x=100$, and with no initial radial velocity.
For the first example of a signal we present the plot for $dt/dx$, where $t$ is the global time coordinate (and not the self-time of the particle), see figure \ref{fig:dtdx}.
The plots are presented for integer $j_L$ values in the range $45\leq j_L\leq 55$.
It is easily seen that an abrupt change of behavior is seen at the value $j_L=50$. No other
change of behavior is seen for other values of $j_L$ in other regions. The plot is focused on the final part of the geodesic, where the change of behavior is observed, at $1.1<x<1.8$.
The second example is the numerical computation of the total falling global time, see figure \ref{fig:total global time}. The points are for trajectories with $j_L$ integer values between $0$ and $100$. The abrupt changing at $j_L=50$ is clear.
The last example is the numerical computation of the total falling self time, see figure \ref{fig:total self time}. The points are for trajectories with $j_L$ integer values between $-25$ and $125$.\footnote{Negative $j_L$ values are for angular momenta in the `other' direction, meaning $\epsilon=-1$.} Again, the abrupt changing at $j_L=50$ is manifest, and it is also manifest that the result is symmetric around the critical point.

In all cases, at the vicinity of the critical value of $j_L=50$, an abrupt change of the behavior of geodesics is seen. Note that $x_0=100$, and so the critical value is $j_L=x_0/2$, which is exactly the angular momentum Fermi-level predicted by the constructed operators of \cite{Berkooz:2006wc}.
Note also, that since the signal appears at $x_0/2$ it suggests that the mismatch found in \cite{Berkooz:2006wc} for the charge-to-angular-momentum scaling \eqref{eq:JQscaling}, in relative to the gravity result \eqref{eq:JQscaling}, is due to a mismatch in $J$ (and not in $Q$).

\section*{Acknowledgments}
 I would like to thank Micha Berkooz for suggesting this work, and for advising throughout. I would also like to thank Zohar Komargodski and Dori Reichmann for many fruitful discussions. Finally, I would like to thank Ofer Aharony for useful discussions and for commenting on an earlier draft.

\end{document}